\documentclass[12pt, epsfig, twocolumn]{aastex6}
\usepackage{graphicx}
\usepackage{color}
\usepackage{epstopdf}

\shorttitle{Local Group distances and publication bias. IV. The
  Galactic Center}
\shortauthors{Richard de Grijs and Giuseppe Bono}

\begin{document}

\title{Clustering of Local Group distances: publication bias or
  correlated measurements? IV. The Galactic Center}

\author{
Richard de Grijs\altaffilmark{1,2} and
Giuseppe Bono\altaffilmark{3,4}
}

\altaffiltext{1} {Kavli Institute for Astronomy \& Astrophysics and
  Department of Astronomy, Peking University, Yi He Yuan Lu 5, Hai
  Dian District, Beijing 100871, China}
\altaffiltext{2} {International Space Science Institute--Beijing, 1
  Nanertiao, Zhongguancun, Hai Dian District, Beijing 100190, China}
\altaffiltext{3} {Dipartimento di Fisica, Universit\`a di Roma Tor
  Vergata, via Della Ricerca Scientifica 1, 00133, Roma, Italy}
\altaffiltext{4} {INAF, Rome Astronomical Observatory, via Frascati
  33, 00040, Monte Porzio Catone, Italy}

\begin{abstract}
Aiming at deriving a statistically well-justified Galactic Center
distance, $R_0$, and reducing any occurrence of publication bias, we
compiled the most comprehensive and most complete database of Galactic
Center distances available to date, containing 273 new or revised
$R_0$ estimates published since records began in October 1918 until
June 2016. We separate our $R_0$ compilation into direct and indirect
distance measurements. The latter include a large body of estimates
that rely on centroid determinations for a range of tracer populations
as well as measurements based on kinematic observations of objects at
the solar circle, combined with a mass and/or rotational model of the
Milky Way. Careful assessment of the Galactic Center distances
resulting from orbital modeling and statistical parallax measurements
in the Galactic nucleus yields our final Galactic Center distance
recommendation of $R_0 = 8.3 \pm 0.2 \mbox{ (statistical)} \pm 0.4
\mbox{ (systematic)}$ kpc. The centroid-based distances are in good
agreement with this recommendation. Neither the direct measurements
nor the post-1990 centroid-based distance determinations suggest that
publication bias may be important. The kinematics-based distance
estimates are affected by significantly larger uncertainties, but they
can be used to constrain the Galaxy's rotation velocity at the solar
Galactocentric distance, $\Theta_0$. Our results imply that the
International Astronomical Union-recommended Galactic Center distance
($R_0^{\rm IAU} = 8.5$ kpc) needs a downward adjustment, while its
$\Theta_0$ recommendation ($\Theta_0 = 220$ km s$^{-1}$) requires a
substantial upward revision.
\end{abstract}


\section{The distance to the Galactic Center}

The distance from the Sun to the Galactic Center, $R_0$, provides the
basic calibration for a wide range of methods used for distance
determination, both on Galactic and extragalactic scales. Calculations
of many physical parameters, including of the distances, masses, and
luminosities of Galactic objects, as well as the Galaxy’s integrated
mass and luminosity, depend directly on $R_0$. In fact, most
luminosity and a large number of mass estimators scale as distance
squared, while masses based on total densities or orbital modeling
scale as distance cubed.

This dependence could involve adoption of a Galactic mass and/or
rotation model, in which case we also need to know the Sun's circular
velocity accurately. As $R_0$ estimates are refined, so are the
estimated distances, masses, and luminosities of numerous Galactic and
extragalactic objects, as well as our best estimates of the rate of
Galactic rotation and the size of the Milky Way. In addition, a highly
accurate direct Galactic Center distance determination would
immediately allow reliable recalibration of the zero points of
numerous secondary distance calibrators, including of Cepheids, RR
Lyrae, and Mira variable stars, which would consequently reinforce the
validity of the extragalactic distance scale (e.g., Olling 2007). In
turn, this would enable better estimates of globular cluster ages, the
Hubble constant, as well as a lower limit to the age of the Universe
(Monelli et al. 2015), and place tighter constraints on a range of
cosmological scenarios (e.g., Reid et al. 2009).

It is no surprise, therefore, that determinations of the Galactic
Center distance have been the subject of many studies ever since the
first attempt by Harlow Shapley in 1914--1918. However, the importance
of determining an accurate $R_0$ value, combined with the large number
of studies undertaken to achieve this goal, have led to speculations
that some degree of publication bias (also known as `observation bias'
or a `bandwagon effect') may have affected subsequent Galactic Center
distance determinations (e.g., Reid 1989, 1993; Nikiforov 2004; Foster
\& Cooper 2010; Malkin 2013a,b; Francis \& Anderson 2014).

In de Grijs et al. (2014, henceforth Paper I), de Grijs \& Bono (2014;
Paper II), and de Grijs \& Bono (2015; Paper III), we embarked on
large-scale data mining of the NASA/Astrophysics Data System (ADS) to
explore whether distance determinations to, respectively the Large
Magellanic Cloud (LMC), the M31 group, and the Small Magellanic Cloud
had been polluted by such bandwagon effects. In this paper, we extend
our series of papers by adding a similar analysis of distance
estimates to the Galactic Center. Our ultimate aim is to provide a
self-consistent distance framework that can serve as a benchmark for
the structure of the Local Group of galaxies (see, e.g., Table 4 in
Paper II, combined with the recommended distance to the Small
Magellanic Cloud from Paper III).

To achieve our aim, in Section 2 we discuss our approach to mining the
NASA/ADS database, eventually resulting in the most complete and most
comprehensive database of Galactic Center distances published,
starting from Shapley's (1918) distance estimate, published in October
1918, until early June 2016. In Section 3, we review three different
types of Galactic Center distance indicators, including direct
distance measurements as well as centroid- and kinematics-based
Galactic Center distance determinations. This is followed in Section 4
by a detailed discussion of the validity of and the uncertainties
affecting the `best' measurements, which we also place in the context
of the most recent progress in the field. Finally, Section 5
summarizes and concludes the paper.

\section{Data Mining}

The distance to the Galactic Center has long been a subject of intense
scrutiny. Since the mid-1970s, it has become standard practice in
meta-analyses of the Galactic Center distance to publish compilations
of previously published values (e.g., Harris 1976; de Vaucouleurs
1983; Kerr \& Lynden-Bell 1986; Reid 1993; Nikiforov 2004; Perryman
2009, his Table 9.1; Malkin 2013a,b). Although the latest such
compilation dates only from 2013, upon close examination it transpired
that Malkin's (2013a,b) compilation of 53 Galactic Center distance
estimates published between 1992 and 2011 is incomplete: our own
perusal of the literature from this same period revealed an additional
32 papers with newly derived or updated Galactic Center distance
estimates.

Since gaps in the data may mask or, alternatively, artificially
suggest the presence of publication bias (for a discussion, see Paper
I), we decided to compile our own database of Galactic Center
distances that is as complete as possible until the present time. We
followed a similar two-pronged approach as employed in Papers I, II,
and III. First, we scanned all 23,516 papers until and including
February 2016 tagged with the `Galactic Center' keyword in the
NASA/ADS for potential new or rederived Galactic Center distance
determinations. At the same time, we carefully followed the reference
trail: where a paper referred to the provenance of the Galactic Center
distance adopted by its authors, we made a note of the original
reference and double checked that the latter was actually included in
our final database.

This approach led to a final database containing a total of 273
Galactic Center distance measurements since Shapley's (1918) first
attempt at determining the centroid of the Galaxtic globular cluster
distribution known at that time. As for Papers I--III, the full
database is available at
http://astro-expat.info/Data/pubbias.html\footnote{A permanent link to
  this page can be found at
  http://web.archive.org/web/20160610121625/http://astro-expat.info/Data/pubbias.html;
  members of the community are encouraged to send us updates or
  missing information, which will be included in the database where
  appropriate.}, where we provide our compilation of Galactic Center
distances both as a function of publication date and by tracer,
supported by full bibliographic information. We compiled the
extinction-corrected distance moduli, as well as their statistical and
systematic uncertainties, if available. Only 25 authors published
their systematic uncertainties separately; in addition, five papers
specified that their published error bars include the systematic
uncertainties. For the remaining Galactic Center distance
measurements, the uncertainties refer to the statistical errors
only. As in Papers I--III, instead of combining individual values
based on different assumptions or input parameters, we have included
all (final) Galactic Center distance measurements published in a given
paper. The range spanned by such alternative values provides a
valuable estimate of the systematic uncertainty inherent to the
distance determined, although we note that these values are often
highly correlated.

\begin{figure*}[ht!]
\plotone{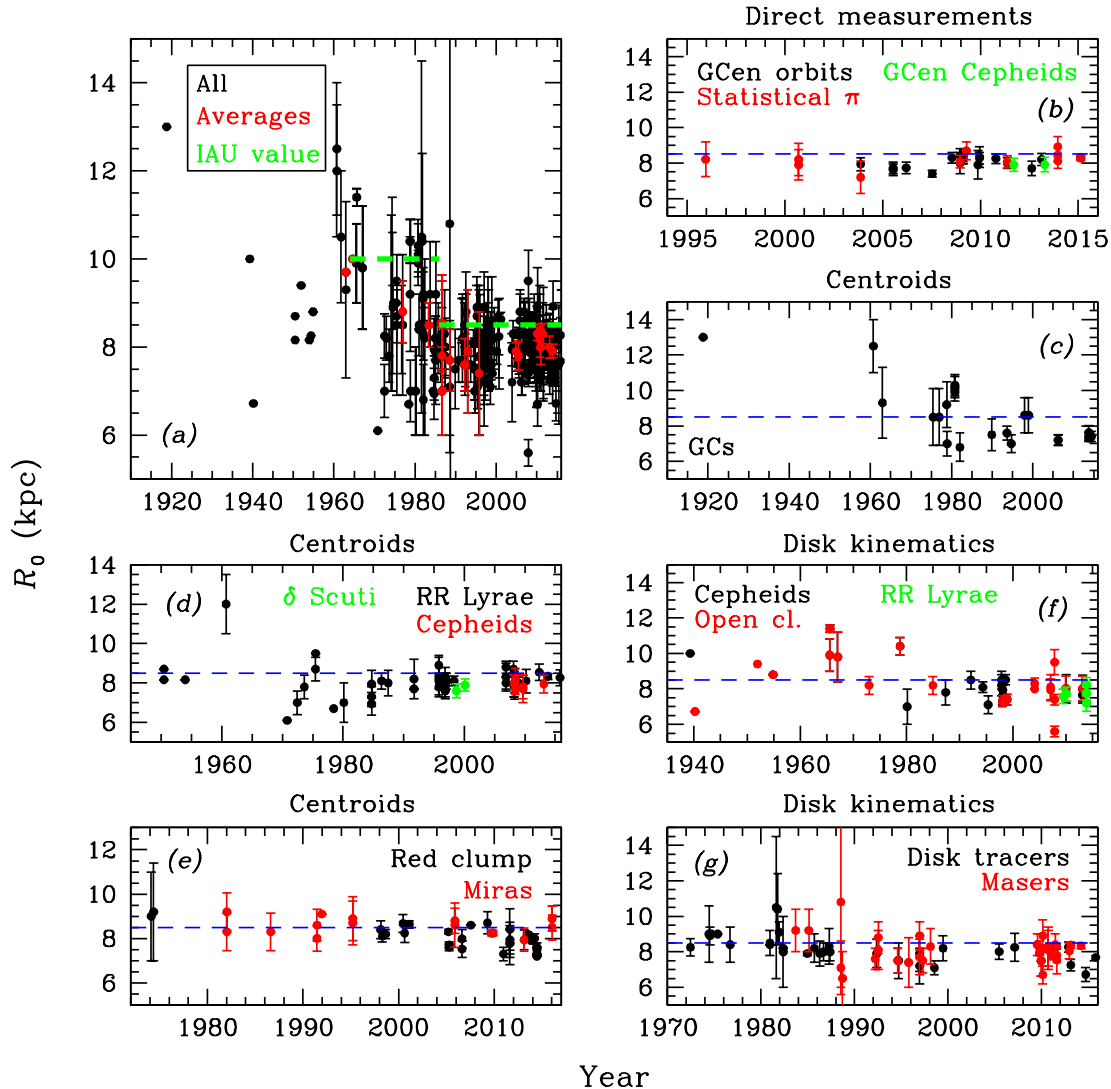}
\caption{Galactic Center distance determinations from the literature
  since records began. (a) All measurements. (b)--(g) Measurements as
  a function of tracer used. The horizontal blue dashed lines indicate
  the IAU-recommended values for $R_0$ at the relevant publication
  dates.}
\label{fig1}
\end{figure*}

Figure \ref{fig1} shows an overview of the data as published in the
original articles, i.e., without having homogenized or recalibrated
the various measurements. We show both the full data set in
Fig. \ref{fig1}a and subsets selected according to their provenance,
i.e., direct measurements (statistical parallaxes, Galactic Center
orbital modeling, Galactic Center Cepheids), centroid determinations
($\delta$ Scuti, RR Lyrae, Cepheids, and Mira variables, red clump
stars), and determinations involving adoption of a kinematic disk
model (Cepheids, RR Lyrae, Miras, open clusters). We have also
included the Galactic Center distance values recommended by the
International Astronomical Union (IAU). At its XII$^{\rm th}$ General
Assembly held in Hamburg (Germany) in 1964, the IAU released its first
formal recommendation for the value of $R_0 = 10$ kpc. This was
subsequently amended to $R_0 = 8.5$ kpc (Kerr \& Lynden-Bell 1986), a
value that has been in force since 1986. Nevertheless, even a casual
comparison of the Galactic Center distances in our database (see
Fig. \ref{fig1}) with the current IAU recommendation reveals that the
latter value is well above the mean of most current and recent
determinations of $R_0$. As such, the community has called for a
further adjustment of the IAU recommendation, but this would need to
be done in tandem with adjustments of the Galactic rotation rate at
the solar circle as well as of the Oort constants, which all depend on
one another.

Our final database is significantly more comprehensive than any
previously published compilation of Galactic Center distance
estimates. We will demonstrate this by comparison of our data set with
those of three benchmark papers, Kerr \& Lynden-Bell (1986), Reid et
al. (1993), and Malkin (2013a). The latter paper was selected for this
comparison since it is the most recent large compilation. As already
addressed earlier in this section, we added 32 papers (containing 42
new Galactic Center distance estimates) to Malkin's (2013a)
compilation of 53 articles published between 1992 May and 2011
July. The review by Reid (1993) has long been used as the `gold
standard' in the field of Galactic Center distance determinations. It
includes 37 $R_0$ values published between 1972 June and 1992 April,
taken from 38 articles. Our compilation, by contrast, includes 20
additional Galactic Center distance estimate, found in 17 papers not
included by Reid (1993). Finally, the current IAU recommendation for
the value of $R_0$ is based on the compilation of 26 estimates by Kerr
\& Lynden-Bell (1986). Again, our own perusal of the literature
published in the period spanning from 1973 August to 1986 May revealed
nine additional Galactic Center distance determinations.

\section{Distances by tracer}

Based on observations taken since 1914, Shapley (1918) used the light
curves and, hence, the period--luminosity relation of Cepheid
variables in 69 Galactic globular clusters to visualize the spatial
distribution of the Galactic globular cluster system. He eventually
extended his analysis to include all 93 Galactic globular clusters
known at that time. He subsequently determined the centroid of the
three-dimensional globular cluster distribution to determine the first
ever distance to the Galactic Center, $13 \le R_0 \le 25$ kpc.

Armed with our current understanding of the structure of the Milky
Way, it is clear that this distance estimate was considerably too
large. In fact, the period--luminosity relation he applied was $\sim$1
mag too faint, while he used Population II Cepheids (W Virginis
stars)---combined with the 25 brightest stars in his seven sample
clusters (since Population II Cepheids were only known in a limited
number of clusters)---instead of the Population I Cepheids he thought
he had observed. Population II Cepheids are some 2 mag fainter than
their Population I counterparts, while it appears that Shapley (1918)
may also have made calibration errors regarding the absolute
magnitudes of the brightest stars and the associated reddening
corrections (Sandage 2004, pp. 301--302), thus offsetting the error
introduced by his use of an incorrect period--luminosity relation to
some extent. As a consequence, Shapley's (1918) distance scale was off
by $\sim$1 mag, corresponding to a factor of $\sim$1.6. Note that the
implicatiom of Shapley's (1918) results based on his approach of using
both Population II Cepheids and his clusters' brightest stars may
suggest that the difference between classical and Population II
Cepheids only applies to a limited number of clusters.

While Shapley's (1918) Galactic Center distance determination now
seems a mere historical curiosity, the fact remains that this was the
first viable `centroid' approach to estimate the Galactic Center's
distance. Nevertheless, it is instructive to compare Shapley's (1918)
result with the most recent $R_0$ determination based on the globular
cluster centroid. Nikiforov \& Smirnova (2013) found that the globular
clusters composing the Galaxy's metal-rich and metal-poor subsystems
separately form bar-like structures that closely resemble the Galactic
bar as a whole. They conclude that only non-axisymmetric models can
provide sufficiently strong constraints on $R_0$. In addition, these
authors found evidence for an extinction component associated with the
Galactic bar which affects the observational incompleteness of
globular clusterss on the far side of the Galactic Center, a selection
effect that must be considered seriously in determining $R_0$.

In this section, we will focus on the `modern' period since 1990 to
distinguish among three different types of Galactic Center distance
determinations in order of decreasing accuracy, i.e., direct
(geometric) methods, centroid-based methods, and distance
determinations based on observations of tracer objects at some
distance from the Galactic Center, combined with a kinematic model of
the Galactic disk. We will also address the uncertainties inherent to
each of the methods applied.

\subsection{Direct measurements}
\label{direct.sec}

The current-best, `direct' (geometric) estimates of $R_0$ are based on
astrometric orbit determinations of the so-called S stars in the
Galactic Center region. Two competing groups are leading efforts in
this field, i.e., Genzel et al. versus Ghez and collaborators: see
Table \ref{GCdirect.tab}. Figure \ref{fig1}b shows the run of these
determinations as a function of publication date (black points). The
latest result from the Genzel group is $R_0 = 8.2 \pm 0.34$ kpc
(Gillessen et al. 2013), while the Ghez group published $R_0 = 8.0 \pm
0.3$ kpc (Yelda et al. 2011; but see Morris et al. 2012 for a rather
surprising downgrade). We will return to a discussion of these direct
measurements in Section \ref{discussion.sec}.

The next best direct Galactic Center estimates come from statistical
parallax measurements of the nuclear star cluster. These measurements
are also included in Table \ref{GCdirect.tab}. Taking the straight
mean and standard deviation of all such measurements gives $R_0 = 8.16
\pm 0.40$ kpc. However, this approach does not do full justice to the
published results. Systematic uncertainties related to the dynamical
cluster models affect the resulting distance determinations. Assuming
uniform, isotropic, and fully phase-mixed systems (e.g., Eisenhauer et
al. 2003; Do et al. 2013) seems to yield systematically smaller $R_0$
values than adopting anisotropic, spherical Jeans models (e.g., Do et
al. 2013). The current-best results are based on joined-up analyses of
both the nuclear star cluster's stellar velocity dispersions and
S-star orbital modeling. The latter lead to $R_0 =
8.46^{+0.42}_{-0.41}$ kpc (Do et al. 2013, combined with Ghez et
al. 2008) and $R_0 = 8.33 \pm 0.11$ kpc (Chatzopoulos et al. 2015,
combined with Gillessen et al. 2009a).

\begin{table*}
\caption{Adopted `direct' distances used in this paper, expressed in
  units of kpc.}
\label{GCdirect.tab}
\begin{center}
{\scriptsize
\tabcolsep 0.5mm
\begin{tabular}{@{}cccll@{}}
\hline \hline
Publ. date & $R_0$ && \multicolumn{1}{c}{Reference} & \multicolumn{1}{c}{Notes} \\
(mm/yyyy) & (kpc) \\
\hline
\multicolumn{5}{c}{Galactic Center orbital modeling}\\
\hline
11/2003 & $7.94 \pm 0.33$ && Eisenhauer et al. (2003)   & S2 only, 1992--2001; systematic uncertainty 0.16 kpc \\
07/2005 & $7.62 \pm 0.32$ && Eisenhauer et al. (2005)   & S2 only, 1992--2004 \\
07/2005 & $7.72 \pm 0.33$ && Eisenhauer et al. (2005)   & S2 only, 1992--2004, excl. data from 2002 \\
03/2006 & $7.73 \pm 0.32$ && Zucker et al. (2006)       & S2 only, 1992--2004; corrected for relativistic effects \\
07/2007 & $7.4  \pm 0.2 $ && Olling (2007; Ghez, priv. commun.) & bias-free `orbital parallax method' (Armstrong et al. 1992) \\
07/2008 & $8.3  \pm 0.3 $ && Ghez et al. (2008)         & S2 only, 1995--2007 \\
12/2008 & $8.0  \pm 0.6 $ && Ghez et al. (2008)         & S2 only, 1995--2007; black hole freely moving \\
12/2008 & $8.4  \pm 0.4 $ && Ghez et al. (2008)         & S2 only, 1995--2007; black hole at rest \\
02/2009 & $8.33 \pm 0.35$ && Gillessen et al. (2009b)   & S stars, 1992--2008; incl. systematic uncertainties \\
02/2009 & $8.40 \pm 0.29$ && Gillessen et al. (2009b)   & S stars excl. S2, 1992--2008; incl. syst. errors \\
12/2009 & $8.28 \pm 0.15$ && Gillessen et al. (2009b)   & S stars, 1992--2008, combined ESO/Keck data sets; syst. unc. 0.29 kpc \\
12/2009 & $8.34 \pm 0.27$ && Gillessen et al. (2009b)   & S2 only, 1992--2008, combined ESO/Keck data sets; syst. unc. 0.52 kpc \\
05/2011 & $8.0  \pm 0.3 $ && Yelda et al. (2011)        & S2 only, 1995--2007; new distortion corrections \\
08/2012 & $7.7  \pm 0.4 $ && Morris et al. (2012)       & S stars, 1995--2011 \\
02/2013 & $8.2  \pm 0.34$ && Gillessen et al. (2013)    & 5 S stars, 1992--2012 \\
\hline
\multicolumn{5}{c}{Nuclear star cluster: Statistical parallaxes}\\
\hline
12 1995 & $8.21 \pm 0.98$ && Huterer et al. (1995)      & 50 M giants \\
09 2000 & $8.2  \pm 0.9 $ && Genzel et al. (2000)       & 104 stars with proper motions; 71 stars with $z$ velocities \\
09 2000 & $7.9  \pm 0.85$ && Genzel et al. (2000)       & Corrected for the effects of a central point mass \\
11 2003 & $7.2  \pm 0.9 $ && Eisenhauer et al. (2003)   & Uniform, isotropic, phase-mixed system \\
12 2008 & $8.07 \pm 0.35$ && Trippe et al. (2008)       & 664 late-type giants \\
05 2011 & $8.07 \pm 0.32$ && Trippe et al. (2011)       & Velocity dispersion; systematic uncertainty 0.13 kpc \\
12 2013 & $8.12^{+0.43}_{-0.41}$ && Do et al. (2013)    & Isotropic velocity distribution \\
12 2013 & $8.92 \pm 0.58$ && Do et al. (2013)           & Anisotropic spherical Jeans models \\
12 2013 & $8.46^{+0.42}_{-0.38}$ && Do et al. (2013)    & Combined with Ghez et al. (2008) \\
02 2015 & $8.27 \pm 0.09$ && Chatzopoulos et al. (2015) & Systematic uncertainty 0.1 kpc \\
02 2015 & $8.33 \pm 0.11$ && Chatzopoulos et al. (2015) & Combined with Gillessen et al. (2009a) \\
\hline \hline
\end{tabular}
}
\end{center}
\end{table*}

\subsection{Galactic Center Cepheids in context}

Matsunaga et al. (2011) used the near-infrared period--luminosity
relation of van Leeuwen et al. (2007), adopting solar metallicity,
calibrated on the basis of Cepheids with parallax-based distances, to
determine the distance modulus to three classical Cepheids in the
Galactic nucleus. These three objects were found within 40 pc
(projected distance) of the central black hole, leading the authors to
conclude that $R_0 = 7.9^{+0.1}_{-0.2}$ kpc (see also Bono et
al. 2013; Matsunaga 2013). The latter value was scaled based on
adoption of $(m-M)_0 = 18.50$ mag for the distance modulus to the
Large Magellanic Cloud. Although this is not strictly speaking a
`direct' Galactic Center distance measurement, we include it here in a
separate subsection rather than in the section where we discuss the
Cepheid centroid given the close proximity of these nuclear Cepheids
to the actual Galactic Center.

The direct measurements discussed in Section \ref{direct.sec} and the
nuclear Cepheid distance of Matsunaga et al. (2011) are indeed fully
consistent with one another, with the Cepheid distance somewhat on the
short side.\footnote{On the other hand, a comparison of Matsunaga et
  al.'s (2011) Galactic Center distance estimate with those based on
  the centroids of the distributions of different Cepheid samples
  (Section \ref{centroids.sec}) suggests that any differences are
  negligible.} If this small distance differential between the nuclear
Cepheids and the actual Galactic Center is real, this may imply that
the former objects are seen in projection onto the Galactic Center. In
fact, Matsunaga et al. (2011) do not claim precise coincidence with
the Galactic Center, but merely that their projected positions are
consistent with their presence in the thin disk-like structure of the
nuclear bulge. These authors point out that their mean distance
estimate suffers from an additional, systematic uncertainty of
approximately 0.3 kpc, which is predominantly driven by the
uncertainties inherent to the extinction law applied (i.e., the
total-to-selective extinction ratio) and by scatter among the
prevailing period--luminosity relations, both corresponding to
uncertainties of 0.09 mag in distance modulus. Matsunaga et al. (2015)
added a fourth nuclear Cepheid to their sample and obtained their
kinematics. They concluded that the velocities of these Cepheids
suggest that the stars orbit within the nuclear stellar disk, i.e.,
within $\sim 200$ pc of the Galactic Center. This is indeed consistent
with their earlier suggestion.

Finally, we note that some authors have suggested that the distances
to different Galactic Center tracers affected by significant
extinction should be adjusted upward, for instance by adopting a
non-standard or variable extinction law toward the Galactic Center
(e.g., Collinge et al. 2006; Vanhollebeke et al. 2009; Pietrukowicz et
al. 2012; Nataf et al. 2013; Nataf et al. 2016; but see Francis \&
Anderson 2014). In addition, the effects of population differences
(ages and metallicities) may cause certain tracer objects---such as
red clump stars---to be redder in the Galactic bulge than their solar
neighborhood equivalents (e.g., Girardi \& Salaris 2001), which thus
implies that significant systematic uncertainties likely remain in the
zero-point calibrations of many `standard' secondary distance tracers.

\subsection{Centroids}
\label{centroids.sec}

Since the pioneering effort by Shapley (1918), many authors have
attempted to determine $R_0$ based on the centroids of the
distributions of a variety of tracers, including globular clusters,
Cepheid, RR Lyrae, and Mira variables, red clumps stars, and even
delta Scuti stars and planetary nebulae.

\begin{figure}[ht!]
\plotone{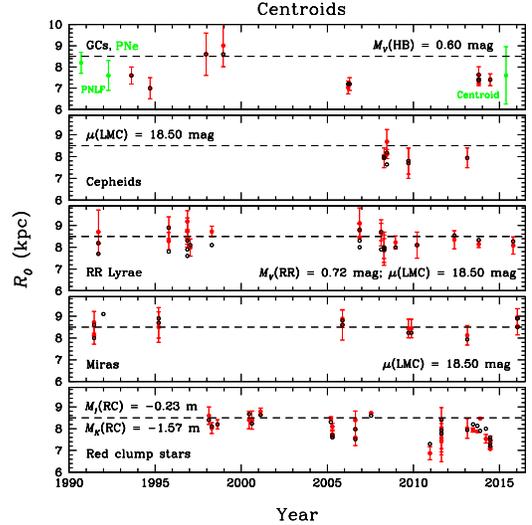}
\caption{Homogenized, post-1990 Galactic Center distance estimates
  based on centroid determinations. The tracers used and the
  calibrations applied are indicated in the individual panels; GCs:
  Globular clusters; PNe: Planetary nebulae; PNLF: PN luminosity
  function; $\mu{\rm (LMC)}$: LMC distance modulus. Where estimates
  have been recalibrated, the original data points are shown as black
  open circles (without error bars for reasons of clarity), while the
  homogenized values appear as red bullets with their corresponding
  error bars. The horizontal dashed lines represent the
  IAU-recommended Galactic Center distance.}
\label{fig2}
\end{figure}

While Fig. \ref{fig1}c--e showed the `raw,' uncorrected centroid data
collected in our database, Fig. \ref{fig2} includes the most recent
(post-1990) Galactic Center measurements based on different centroid
tracers as a function of publication date. We have attempted to
homogenize the distance calibration where appropriate: the original
data points as listed in the online database are shown as black open
circles; the corrected measurements are shown as red bullets with
error bars.

The calibrations applied are included in the individual panels. We
used a common horizontal-branch magnitude of $M_V {\rm (HB)} = 0.60$
mag for globular clusters, $M_V {\rm (RR)} = 0.72$ mag for RR Lyrae
stars, and $I$- and $K$-band red clump absolute magnitudes from {\sl
  Hipparcos} (Stanek \& Garnavich 1998; see also Laney et al. 2012)
for the red clump stars, i.e., $M_I^{\rm RC} = -0.23$ mag and
$M_K^{\rm RC} = -1.57$ mag, with $M_K^{\rm RC} = M_{K_{\rm s}}^{\rm
  RC} + 0.044$ mag (Grocholski \& Sarajedini 2002). In addition, we
adopted a homogenized distance scale based on $(m-M)_0^{\rm LMC} =
18.50$ mag, thus also ensuring internal consistency with the
recommended distance benchmarks derived in Papers I--III.

\begin{table*}
\caption{Adopted, homogenized centroid-based distances used in this
  paper. Entries in {\bf bold font} are newly updated (homogenized)
  distance estimates.}
\label{GCcentroids.tab}
\begin{center}
\tabcolsep 0.5mm
{\scriptsize
\begin{tabular}{@{}cclll@{}}
\hline \hline
Publ. date & $R_0$ & \multicolumn{1}{c}{Reference} & \multicolumn{1}{c}{Original Calibration} & \multicolumn{1}{c}{Notes$^a$} \\
(mm/yyyy) & (kpc) \\
\hline
\multicolumn{5}{c}{Globular clusters; $\langle {\rm [Fe/H]}^{\rm GCs} \rangle = -1.5$ dex}\\
\hline
08/1993 & $7.6  \pm 0.4 $ & Maciel (1993)                & $M_V{\rm (HB)} = 0.6$ mag & 46 GCs, ${\rm [Fe/H]} > -1.2$ dex \\
09/1994 & $7    \pm 0.5 $ & Rastorguev et al. (1994)     & $M_V{\rm (RR)} = 0.6$ mag & $\cdots$ \\
00/1998 & $8.6  \pm 1.0 $ & Surdin \& Feoktistov (unpubl.) & $\cdots$                  & quoted in Surdin (1999) as `submitted' \\
00/1999 & ${\bf 9.0} \pm 1.0 $ & Surdin (1999)           & $M_V{\rm (HB)} = 0.20 {\rm [Fe/H]} + 1.00$ mag & 126 GCs  \\
03/2006 & ${\bf 7.0} \pm 0.3 $ & Bonatto et al. (2009)   & Calibration from Bica et al. (2006)            & $\cdots$ \\
04/2006 & $7.2  \pm 0.3 $ & Bica et al. (2006)           & $M_V{\rm (HB)} = 0.16 {\rm [Fe/H]} + 0.84$ mag & 116 GCs \\
10/2013 & $7.63 \pm 0.38$ & Nikiforov \& Smirnova (2013) & $M_V{\rm (HB)} = 0.16 {\rm [Fe/H]} + 0.84$ mag & all methods \\
10/2013 & $7.42 \pm 0.23$ & Nikiforov \& Smirnova (2013) & $M_V{\rm (HB)} = 0.16 {\rm [Fe/H]} + 0.84$ mag & all spatial methods\\
10/2013 & $7.36 \pm 0.24$ & Nikiforov \& Smirnova (2013) & $M_V{\rm (HB)} = 0.16 {\rm [Fe/H]} + 0.84$ mag & Shapley's method and related spatial methods \\
06/2014 & $7.4  \pm 0.2 $ & Francis \& Anderson (2014)   & $M_V{\rm (RR)} = 1.067 + 0.502 {\rm [M/H]}$    & 154 GCs, syst. unc. 0.2 kpc \\
        &                 &                              & $+ 0.108 {\rm [M/H]}^2$ mag \\
\hline
\multicolumn{5}{c}{Cepheids}\\
\hline
04/2008 & $7.94 \pm 0.37$ & Groenewegen et al. (2008)    & $(m-M)_0^{\rm LMC} = 18.50$ mag & Cepheids and RR Lyrae; syst. unc. 0.26 kpc \\
04/2008 & $7.99 \pm 0.09$ & Groenewegen et al. (2008)    & $(m-M)_0^{\rm LMC} = 18.50$ mag & Type II Cepheids \\
06/2008 & ${\bf 8.11} \pm 0.21$ & Feast et al. (2008)    & $(m-M)_0^{\rm LMC} = 18.37$ mag & Type II Cepheids \\
06/2008 & ${\bf 8.69} \pm 0.56$ & Feast et al. (2008)    & $(m-M)_0^{\rm LMC} = 18.37$ mag & Type II Cepheids, incl. $\kappa$ Pav \\
09/2009 & $7.8  \pm 0.6 $ & Majaess et al. (2009)        & $(m-M)_0^{\rm LMC} = 18.50$ mag & Type II Cepheids \\
09/2009 & $7.7  \pm 0.7 $ & Majaess et al. (2009)        & $(m-M)_0^{\rm LMC} = 18.50$ mag & Type II Cepheids + bulge model \\
02/2013 & $7.94 \pm 0.37$ & Matsunaga (2013)             & $(m-M)_0^{\rm LMC} = 18.50$ mag & 100 Miras, Cepheids, RC stars; syst. unc. 0.26 kpc \\
\hline
\multicolumn{5}{c}{RR Lyrae; $\langle {\rm [Fe/H]}^{\rm RR} \rangle = -1.0$ dex}\\
\hline
09/1991 & ${\bf 8.7} \pm 1.0 $ & Walker \& Terndrup (1991) & $M_V{\rm (RR)} = 0.85$ mag    & $N = 44$, Baade's Window; $R_V = 3.1$ \\
09/1991 & ${\bf 8.2} \pm 0.  $ & Walker \& Terndrup (1991) & $M_V{\rm (RR)} = 0.85$ mag    & $N = 44$, Baade's Window; $R_V = 3.35$, \\ 
        &                 &                              &                                 & Carney et al. (1992) distance scale \\
10/1995 & ${\bf 8.28} \pm 0.40$ & Carney et al. (1995)   & $M_V{\rm (RR)} = 0.85$ mag      & $N = 58$, Baade's Window \\
10/1995 & $8.9  \pm 0.5 $ & Carney et al. (1995)         & $(m-M)_0^{\rm LMC} = 18.50$ mag & $N = 58$, Baade's Window (but see Feast 1997) \\
11/1996 & ${\bf 8.4} \pm 0.4 $ & Layden et al. (1996)    & $(m-M)_0^{\rm LMC} = 18.28$ mag & Baade's Window \\
        &                 &                              & $M_V{\rm (RR)} = 0.15 {\rm [Fe/H]} + 0.95$ mag \\
11/1996 & ${\bf 9.2} \pm 0.5 $ & Layden et al. (1996)    & $(m-M)_0^{\rm LMC} = 18.28$ mag & Baade's Window \\
        &                 &                              & $M_V{\rm (RR)} = 0.60$ mag \\
11/1996 & ${\bf 8.7} \pm 0.6 $ & Layden et al. (1996)    & Combined approach               & `best' value \\
01/1997 & ${\bf 8.0} \pm 0.4 $ & Feast (1997)            & $(m-M)_0^{\rm LMC} = 18.53$ mag & $N = 18$ \\
        &                 &                              & $M_V{\rm (RR)} = 0.37 {\rm [Fe/H]} + 1.13$ mag \\
04/1998 & ${\bf 8.7} \pm 0.25$ & Udalski (1998)          & $M_V{\rm (RR)} = 0.88$ mag      & $N = 73$ \\
11/2006 & $8.8  \pm 0.3 $ & Collinge et al. (2006)       & $M_V{\rm (RR)} = 0.72$ mag      & $N = 159$, Baade's Window \\
11/2006 & ${\bf 8.8} \pm 0.3 $ & Collinge et al. (2006)  & $M_V{\rm (RR)} = 0.92$ mag      & $N = 159$, Baade's Window \\
11/2006 & ${\bf 9.1} \pm 0.7 $ & Collinge et al. (2006)  & Combined approach               & $N = 159$, Baade's Window \\
02/2008 & $8.7  \pm 0.4 $ & Kunder et al. (2008)         & $M_V{\rm (RR)} = 0.72$ mag      & $\cdots$ \\
02/2008 & ${\bf 8.7} \pm 0.6 $ & Kunder et al. (2008)    & $M_V{\rm (RR)} = 0.92$ mag      & $\cdots$ \\
04/2008 & $7.94 \pm 0.37$ & Groenewegen et al. (2008)    & $(m-M)_0^{\rm LMC} = 18.50$ mag & Cepheids+RR Lyrae \\
04/2008 & $7.87 \pm 0.64$ & Groenewegen et al. (2008)    & $(m-M)_0^{\rm LMC} = 18.50$ mag & $N = 37$, Sollima et al. (2006) PLC calibration \\
04/2008 & $8.0  \pm 0.7 $ & Groenewegen et al. (2008)    & $(m-M)_0^{\rm LMC} = 18.50$ mag & $N = 37$, Sollima et al. (2006) PLR calibration \\
12/2008 & ${\bf 8.2} \pm 0.3 $ & Kunder \& Chaboyer (2008) & $M_V{\rm (RR)} = 1.19 + 0.5 {\rm [Fe/H]}$ & $N = 2690$, bulge \\
        &                 &                              & $+ 0.09 {\rm [Fe/H]}^2$ mag \\
03/2010 & $8.1  \pm 0.6 $ & Majaess (2010)               & $(m-M)_0^{\rm LMC} = 18.50$ mag & $\cdots$ \\
05/2012 & ${\bf 8.35} \pm 0.42$ & Pietrukowicz et al. (2012) & $M_V{\rm (RR)} = 2.288 + 0.882 \log Z$    & $\log Z \equiv {\rm [Fe/H]} - 1.765$ dex \\
        &                 &                              & $+ 0.108 (\log Z)^2$ mag \\
10/2013 & ${\bf 8.14} \pm 0.05$ & D\'ek\'any et al. (2013)   & Catelan et al. (2004) calibration & Systematic uncertainty 0.14 kpc \\
10/2015 & ${\bf 8.08} \pm 0.01$ & Pietrukowicz et al. (2014) & Catelan et al. (2004) calibration & Systematic uncertainty 0.40 kpc \\
\hline \hline
\end{tabular}
}
\end{center}
\end{table*}

\addtocounter{table}{-1}
\begin{table*}
\caption{(Continued)}
\begin{center}
\tabcolsep 0.5mm
{\scriptsize
\begin{tabular}{@{}cclll@{}}
\hline \hline
Publ. date & $R_0$ & \multicolumn{1}{c}{Reference} & \multicolumn{1}{c}{Original Calibration} & \multicolumn{1}{c}{Notes} \\
(mm/yyyy) & (kpc) \\
\hline
\multicolumn{5}{c}{Miras}\\
\hline
06/1991 & ${\bf 8.7} \pm 0.5 $ & Whitelock et al. (1991)    & $(m-M)_0^{\rm LMC} = 18.47$ mag & $\cdots$ \\
06/1991 & ${\bf 8.1} \pm 0.5 $ & Whitelock et al. (1991)    & $(m-M)_0^{\rm LMC} = 18.47$ mag & $\cdots$ \\
00/1992 & $9.1               $ & Whitelock (1992)           & $\cdots$                        & $\cdots$ \\
03/1995 & ${\bf 8.5} \pm 0.7 $ & Glass et al. (1995)        & $(m-M)_0^{\rm LMC} = 18.55$ mag & $\cdots$ \\
03/1995 & ${\bf 8.7} \pm 0.7 $ & Glass et al. (1995)        & $(m-M)_0^{\rm LMC} = 18.55$ mag & If located in the Galactic bar \\
11/2005 & $8.6  \pm 0.7 $ & Groenewegen \& Blommaert (2005) & $(m-M)_0^{\rm LMC} = 18.50$ mag & $N = 2691$ \\
11/2005 & ${\bf 8.9} \pm 0.4 $ & Groenewegen \& Blommaert (2005) & $(m-M)_0^{\rm LMC} = 18.48$ mag & $N = 2691$; Feast (2004) calibration \\
09/2009 & ${\bf 8.43} \pm 0.08$ & Matsunaga et al. (2009)   & $(m-M)_0^{\rm LMC} = 18.45$ mag & Systematic uncertainty 0.42 kpc \\
11/2009 & ${\bf 8.43} \pm 0.08$ & Matsunaga et al. (2009)   & $(m-M)_0^{\rm LMC} = 18.45$ mag & Systematic uncertainty 0.42 kpc \\
02/2013 & ${\bf 8.13} \pm 0.37$ & Matsunaga (2013)          & $(m-M)_0^{\rm LMC} = 18.45$ mag & Cepheids, Miras, RC stars; syst. unc. 0.26 kpc \\
01/2016 & $8.9  \pm 0.4 $ & Catchpole et al. (2016)         & $(m-M)_0^{\rm LMC} = 18.49$ mag & $2.1 \le \log P {\rm [days]} < 2.6$ \\
01/2016 & $8.5  \pm 0.4 $ & Catchpole et al. (2016)         & $(m-M)_0^{\rm LMC} = 18.49$ mag & $2.6 \le \log P {\rm [days]} < 2.7$ \\
01/2016 & $8.9  \pm 0.4 $ & Catchpole et al. (2016)         & $(m-M)_0^{\rm LMC} = 18.49$ mag & Combined approach \\
\hline
\multicolumn{5}{c}{Red clump stars}\\
\hline
02/1998 & ${\bf 8.6} \pm 0.4 $ & Paczy\'nski \& Stanek (1998) & $M_I^0 = -0.28$ mag             & $\cdots$ \\
04/1998 & ${\bf 8.0} \pm 0.25$ & Udalski (1998)             & $M_I= (0.09 \pm 0.03) {\rm [Fe/H]}^{\rm RC}$ & ${\rm [Fe/H]}^{\rm RC} = +0.2$ dex; 73 RRab Lyrae, RC stars \\
        &                 &                                 & $-(0.23 \pm 0.03)$ mag \\
08/1998 & $8.2  \pm 0.15$ & Stanek \& Garnavich (1998)      & $M_I = -0.23$ mag               & $\cdots$ \\
06/2000 & ${\bf 8.40} \pm 0.4 $ & Stanek et al. (2000)      & $M_I^0 = -0.16$ mag             & $\cdots$ \\
08/2000 & ${\bf 8.39} \pm 0.42$ & Alves (2000)              & $M_K = -1.61 \pm 0.03$ mag      & $\cdots$ \\
02/2001 & ${\bf 8.79} \pm 0.16$ & Gould et al. (2001)       & $M_K = -1.61$ mag               & Updated zero points \\
03/2005 & ${\bf 8.5} \pm 0.1 $  & Nishiyama et al. (2005)   & $M_K = -1.61$ mag               & $\cdots$ \\
04/2005 & ${\bf 8.1} \pm 0.15$  & Babusiaux \& Gilmore (2005) & $M_K = -1.68$ mag             & $\cdots$ \\
04/2005 & ${\bf 7.7} \pm 0.15$  & Babusiaux \& Gilmore (2005) & $M_K = -1.61$ mag             & $\cdots$ \\
08/2006 & ${\bf 7.59} \pm 0.10$ & Nishiyama et al. (2006)   & $M_K = -1.59$ mag               & $\cdots$ \\
08/2006 & ${\bf 8.40} \pm 0.42$ & Nishiyama et al. (2006)   & $M_K = -1.68$ mag               & Recalibration of Alves (2000) \\
07/2007 & ${\bf 8.7}          $ & Rattenbury et al. (2007)  & $M_I^0 = -0.26$ mag             & Upper limit \\
12/2010 & ${\bf 6.9} \pm 0.3 $  & McWilliam \& Zoccali (2010) & $M_K = -1.44$ mag             & $\cdots$ \\
08/2011 & ${\bf 7.43} \pm 0.63$ & Fritz et al. (2011)       & $M_{K_{\rm s}} = -1.47$ mag     & $K_{\rm s}$ \\
08/2011 & ${\bf 7.43} \pm 0.95$ & Fritz et al. (2011)       & $M_{K_{\rm s}} = -1.47$ mag     & $H$ \\
08/2011 & ${\bf 8.03} \pm 0.94$ & Fritz et al. (2011)       & $M_{K_{\rm s}} = -1.47$ mag     & $L'$ \\
08/2011 & ${\bf 7.58} \pm 0.65$ & Fritz et al. (2011)       & $M_{K_{\rm s}} = -1.47$ mag     & mean \\
02/2013 & ${\bf 8.01} \pm 0.37$ & Matsunaga (2013)          & $M_K = -1.59$ mag               & Cepheids, Miras, RC stars \\
06/2013 & ${\bf 7.94} \pm 0.1 $ & Nataf et al. (2013)       & $M_{K_{\rm s}} = -1.50$ mag     & $\cdots$ \\
09/2013 & ${\bf 7.87}         $ & Cao et al. (2013)         & $M_{K_{\rm s}} = -1.50$ mag     & $\cdots$ \\
11/2013 & ${\bf 8.5}          $ & Wegg \& Gerhard (2013)    & $M_K = -1.72$ mag               & $\cdots$ \\
03/2014 & ${\bf 7.5}  \pm 0.2 $ & Gardner et al. (2014)     & $M_K = -1.44$ mag               & $\cdots$ \\
06/2014 & ${\bf 7.4} \pm 0.3 $ & Francis \& Anderson (2014) & $M_K = -1.53 \pm 0.01$ mag      & {\sl Hipparcos} recalibration \\
        &                 &                                 & $M_I = -0.24 \pm 0.01$ mag \\
06/2014 & ${\bf 7.5}    $ & Francis \& Anderson (2014)      & $M_K = -1.53 \pm 0.01$ mag      & Recalibration of Alves (2000) \\
        &                 &                                 & $M_I = -0.24 \pm 0.01$ mag \\
06/2014 & ${\bf 7.2}    $ & Francis \& Anderson (2014)      & $M_K = -1.53 \pm 0.01$ mag      & Recalibration of Babusiaux \& Gilmore (2005) \\
        &                 &                                 & $M_I = -0.24 \pm 0.01$ mag \\
06/2014 & ${\bf 7.1}    $ & Francis \& Anderson (2014)      & $M_K = -1.53 \pm 0.01$ mag      & Recalibration of Nishiyama et al. (2006) \\
        &                 &                                 & $M_I = -0.24 \pm 0.01$ mag \\
\hline \hline
\end{tabular}
}
\end{center}
\flushleft
$^a$ Abbreviations used: GCs, globular clusters; $N$, number of
objects; $P$, period; PLC, period--luminosity--color relation; PLR,
period--luminosity relation; RC, red clump; syst. unc., systematic
uncertainty.
\end{table*}

The left-hand side of Table \ref{tracers.tab} provides an overview of
the mean Galactic Center distances and their standard deviations based
on published centroid determinations since 1990. It is readily
apparent that the RR Lyrae and Mira centroids yield systematically
larger distances than the other tracers. We checked whether this might
be a signature of publication bias by recalculating the RR Lyrae
statistics for Galactic Center distances published since 2000. Their
(mean, $\sigma$) combinations are (8.21, 0.31) kpc and (8.31, 0.39)
kpc before and after correction, respectively.

Determination of $R_0$ based on the centroid approach is significantly
more uncertain than using direct geometry. Uncertainties hindering
accurate Galactic Center distance estimation include ambiguities
associated with the prevailing extinction law, a preference for
smaller values of $R_0$ because of sampling biases (where extinction
causes tracers on the near side of the bulge to have a greater chance
of inclusion than their counterparts on the far side), and
difficulties in converting the mean distance to one's tracer sample
into a value of $R_0$.

\begin{table*}
\caption{Mean original and homogenized Galactic Center distances and
  their standard deviations ($\sigma$) based on published post-1990
  estimates. All values are expressed in units of kpc; note that the
  numbers of data points contributing to the `before' and `after'
  values for a given tracer are not necessarily the same.}
\label{tracers.tab}
\begin{center}
\tabcolsep 0.5mm
\begin{tabular}{@{}lcccccclccccc@{}}
\hline \hline
\multicolumn{6}{c}{Centroids} && \multicolumn{6}{c}{Kinematics} \\
\cline{1-6}\cline{8-13}
& \multicolumn{2}{c}{Before correction} && \multicolumn{2}{c}{After correction} &&& \multicolumn{2}{c}{Before correction} && \multicolumn{2}{c}{After correction} \\
\cline{2-3}\cline{5-6}\cline{9-10}\cline{12-13}
\multicolumn{1}{c}{Tracer} & Mean & $\sigma$ && Mean & $\sigma$ && \multicolumn{1}{c}{Tracer} & Mean & $\sigma$ && Mean & $\sigma$ \\
\hline
Globular clusters & 7.63 & 0.66 && 7.60 & 0.56 && Cepheids+RR Lyrae & 7.88 & 0.44 && 8.15 & 0.43 \\
Cepheids          & 8.02 & 0.32 && 7.88 & 0.19 && Open clusters     & 7.73 & 0.98 && 7.76 & 1.07 \\
RR Lyrae          & 8.15 & 0.34 && 8.45 & 0.39 && Disk tracers      & 7.68 & 0.49 && $\cdots$ \\
Miras             & 8.57 & 0.35 && 8.58 & 0.27 && Masers            & 8.01 & 0.53 && $\cdots$ \\
Red clump stars   & 7.91 & 0.53 && 7.96 & 0.42 \\
\hline \hline
\end{tabular}
\end{center}
\end{table*}

\subsection{Kinematic tracers}

\begin{figure}[ht!]
\plotone{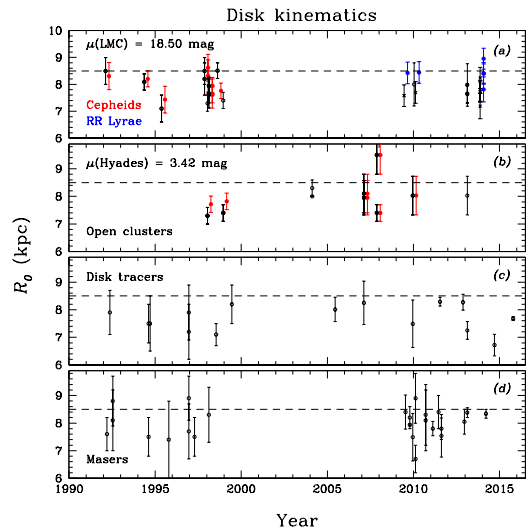}
\caption{Post-1990 Galactic Center distance estimates based on
  kinematic disk models for a variety of tracers, as indicated in the
  individual panels. Where possible, we have homogenized the Galactic
  Center distances based on RR Lyrae and Cepheid variables, as well as
  those based on open cluster distances. The original data points are
  shown in black, with their homogenized counterparts represented in
  color and slightly offset along the horizontal axis. In the top
  panel, the Cepheid-based original values are indicated by open
  circles; those based on RR Lyrae distances are shown as crosses. The
  horizontal dashed lines represent the IAU-recommended Galactic
  Center distance.}
\label{fig3}
\end{figure}

A large number of Galactic Center distance determinations base their
estimates on the distances to a variety of tracer populations at or
near the solar circle, combined with a kinematic disk (mass) model of
the Milky Way galaxy. Suitable tracers include Cepheid and RR Lyrae
variable stars, open clusters, and maser sources, as well as the
distributions of H{\sc i} and CO. Figure \ref{fig3} shows the relevant
measurements published since 1990. As before, we have attempted to
homogenize these estimates where possible, particularly the Cepheid-,
RR Lyrae-, and open cluster-based distances. The latter are
benchmarked with respect to the distance to the Hyades, $(m-M)_0^{\rm
  Hya} = 3.42$ mag, which also ensures internal consistency with the
LMC distance modulus adopted in this series of papers. The mean $R_0$
values and their associated standard deviations based on kinematic
disk modeling are included on the right-hand side of Table
\ref{tracers.tab}.

Note, however, that this method of Galactic Center distance
determination suffers from larger uncertainties than the centroid
approach, given the larger number of assumptions one has to make. In
addition to the need for an accurate Milky Way mass model (e.g., Reid
et al. 2009; McMillan 2011), the basic, underlying assumption on which
the accuracy of this methodology hinges is that the tracers are
stationary with respect to the local standard of rest (at the solar
position) and in circular motion at the solar circle. Since both of
these assumptions are usually imprecise, good results cannot be
expected from individual objects.

\begin{table*}
\caption{Adopted, homogenized kinematics-based distances used in this
  paper. Entries in {\bf bold font} are newly updated (homogenized)
  distance estimates.}
\label{GCkinematics.tab}
\begin{center}
\tabcolsep 0.5mm
{\scriptsize
\begin{tabular}{@{}cclccl@{}}
\hline \hline
Publ. date & $R_0$ & \multicolumn{1}{c}{Reference} & \multicolumn{1}{c}{Original Calibration} & \multicolumn{1}{c}{$\Theta_0$} & \multicolumn{1}{c}{Notes$^a$} \\
(mm/yyyy) & (kpc) & & & (km s$^{-1}$) \\
\hline
\multicolumn{6}{c}{Cepheids}\\
\hline
02/1992 & ${\bf 8.3}  \pm 0.5 $ & Caldwell et al. (1992) & $(m-M)_0^{\rm LMC} = 18.55$ mag & $248 \pm 16$ & $N = 212$ \\
05/1994 & ${\bf 8.20} \pm 0.30$ & Pont et al. (1994)     & $(m-M)_0^{\rm LMC} = 18.47$ mag & $257 \pm  7$ & $\cdots$ \\
05/1995 & ${\bf 7.4}  \pm 0.5 $ & Dambis et al. (1995)      & $(m-M)_0^{\rm Hya} =  3.30$ mag & $\cdots$     & $\cdots$ \\
11/1997 & ${\bf 8.6}  \pm 0.5 $ & Feast \& Whitelock (1997) & $(m-M)_0^{\rm LMC} = 18.47$ mag & $\cdots$     & $R_0$ from Pont et al. (1994); new PLR \\
11/1997 & ${\bf 8.3}  \pm 0.6 $ & Feast \& Whitelock (1997) & $(m-M)_0^{\rm LMC} = 18.47$ mag & $\cdots$     & $R_0$ from Metzger et al. (1998); new PLR \\
01/1998 & ${\bf 7.7}  \pm 0.3 $ & Glushkova et al. (1998)   & $(m-M)_0^{\rm Hya} =  3.30$ mag & $200 \pm 15$ & $N = 202$: open clusters, Cepheids, RSGs \\
02/1998 & $7.66 \pm 0.32$       & Metzger et al. (1998)     & $(m-M)_0^{\rm LMC} = 18.50$ mag & $237 \pm 12$ & Systematic uncertainty 0.44 kpc \\
02/1998 & $7.95 \pm 0.31$       & Metzger et al. (1998)     & $(m-M)_0^{\rm LMC} = 18.50$ mag & $237 \pm 12$ & $N = 7$, solar circle; no orbital ellipticity \\
02/1998 & $7.61 \pm 0.30$       & Metzger et al. (1998)     & $(m-M)_0^{\rm LMC} = 18.50$ mag & $237 \pm 12$ & $N = 7$, solar circle; 4\% orbital ellipticity \\
08/1998 & ${\bf 7.76} \pm 0.29$ & Feast et al. (1998)       & $(m-M)_0^{\rm LMC} = 18.70$ mag & $\cdots$     & $N = 266$ \\
00/1999 & ${\bf 7.8}  \pm 0.3 $ & Glushkova et al. (1999)   & $(m-M)_0^{\rm Hya} =  3.30$ mag & $204 \pm 15$ & $N = 202$: open clusters, Cepheids, RSGs; use caution$^b$ \\
01/2010 & $8.0  \pm 0.8 $       & Shen \& Zhang (2010)      &                                 & $239 \pm 12$ & $\cdots$ \\
02/2013 & $7.66 \pm 0.36$       & Bobylev (2013)            & $(m-M)_0^{\rm Hya} =  3.42$ mag & $267 \pm 17$ & $N = 14$; $P > 5$ days \\
02/2013 & $7.64 \pm 0.32$       & Bobylev (2013)            & $(m-M)_0^{\rm Hya} =  3.42$ mag & $217 \pm 11$ & $N = 18$; proper motions \\
02/2013 & $7.98 \pm 0.79$       & Zhu \& Shen (2013)        & $(m-M)_0^{\rm Hya} =  3.42$ mag & $239 \pm 23$ & $N = 215$ \\
\hline
\multicolumn{6}{c}{RR Lyrae}\\
\hline
06/2009 & ${\bf 8.43} \pm 0.40$ & Dambis (2009)             & $(m-M)_0^{\rm LMC} = 18.27$ mag & $229 \pm 12$ & $\cdots$ \\
02/2010 & ${\bf 8.4}  \pm 0.4 $ & Dambis (2010)             & $(m-M)_0^{\rm LMC} = 18.30$ mag & $195 \pm  5$ & $\cdots$ \\
11/2013 & ${\bf 8.41} \pm 0.40$ & Dambis et al. (2013)      & $(m-M)_0^{\rm LMC} = 18.32$ mag & $234 \pm 10$ & Recalibration of Carney et al. (1995) \\
11/2013 & ${\bf 7.81} \pm 0.40$ & Dambis et al. (2013)      & $(m-M)_0^{\rm LMC} = 18.32$ mag & $234 \pm 10$ & Recalibration of Groenewegen et al. (2008) \\
11/2013 & ${\bf 8.95} \pm 0.39$ & Dambis et al. (2013)      & $(m-M)_0^{\rm LMC} = 18.32$ mag & $234 \pm 10$ & Recalibration of Collinge et al. (2006) \\
11/2013 & ${\bf 8.40} \pm 0.36$ & Dambis et al. (2013)      & $(m-M)_0^{\rm LMC} = 18.32$ mag & $234 \pm 10$ & Average value \\
\hline
\multicolumn{6}{c}{Open clusters}\\
\hline
01/1998 & ${\bf 7.7} \pm 0.3 $ & Glushkova et al. (1998) & $(m-M)_0^{\rm Hya} =  3.30$ mag & $200 \pm 15$ & $N = 202$: open clusters, Cepheids, RSGs \\
00/1999 & ${\bf 7.8} \pm 0.3 $ & Glushkova et al. (1999) & $(m-M)_0^{\rm Hya} =  3.30$ mag & $204 \pm 15$ & $N = 202$: open clusters, Cepheids, RSGs; use caution$^b$ \\
02/2004 & $8.3  \pm 0.3 $ & Gerasimenko (2004)        & $\cdots$                        & $\cdots$     & $N = 146$; Weaver's method \\
02/2004 & $8.0          $ & Gerasimenko (2004)        & $\cdots$                        & $\cdots$     & $N = 34$; Fokker's method \\
02/2007 & $7.95 \pm 0.62$ & Shen \& Zhu (2007)        & $(m-M)_0^{\rm Hya} =  3.42$ mag & $244 \pm 14$ & $N = 270$ \\
02/2007 & $8.1  \pm 0.7 $ & Shen \& Zhu (2007)        & $(m-M)_0^{\rm Hya} =  3.42$ mag & $244 \pm 14$ & Mean value (open clusters, OB stars) \\
11/2007 & $7.4  \pm 0.3 $ & Bobylev et al. (2007)     & $(m-M)_0^{\rm Hya} =  3.42$ mag & $195 \pm  7$ & $N = 375$ \\
11/2007 & $9.5  \pm 0.7 $ & Bobylev et al. (2007)     & $(m-M)_0^{\rm Hya} =  3.42$ mag & $195 \pm  7$ & Young open clusters ($< 50$ Myr) \\
11/2007 & $5.6  \pm 0.3 $ & Bobylev et al. (2007)     & $(m-M)_0^{\rm Hya} =  3.42$ mag & $195 \pm  7$ & Old open clusters ($> 50$ Myr) \\
12/2009 & $8.03 \pm 0.70$ & Zhu (2009)                & $(m-M)_0^{\rm Hya} =  3.42$ mag & $235 \pm 10$ & $N = 301$ \\
02/2013 & $8.03 \pm 0.70$ & Zhu \& Shen (2013)        & $(m-M)_0^{\rm Hya} =  3.42$ mag & $260 \pm 15$ & $N = 301$ \\
\hline \hline
\end{tabular}
}
\end{center}
\flushleft 
$^a$ New abbreviation used: RSGs, red supergiant stars.\\ 
$^b$ Note that Glushkova et al. (1998) and Glushkova et al. (1999) are
identical peer-reviewed articles, although this duplicate publication
is not formally acknowledged. Nevertheless, the Galactic Center
distance estimates differ by 0.1 kpc between both papers, which
remains unexplained.
\end{table*}

It is not obvious how one could self-consistently homogenize the
Galactic Center distance estimates based on other kinematic disk
tracers. We considered using the ratio of the rotational velocities at
the solar circle, $\Theta_0$, and $R_0$ to transfer all these
measurements onto a common scale. McMillan \& Binney (2010) carefully
re-evaluated a number of current models describing solar rotation
about the Galactic Center. Using data of 18 masers in high-mass
star-forming regions at the solar circle from Reid et al. (2009), they
found that the best-fitting models yield Galactic Center distances in
the range from $R_0 = 6.7 \pm 0.5$ kpc to $R_0 = 8.9 \pm 0.9$, and
$\Theta_0$ from $200 \pm 20$ km s$^{-1}$ to $279 \pm 33$ km s$^{-1}$.
Despite these large ranges in $R_0$ and $\Theta_0$, which are largely
driven by one's choice of Galactic rotation curve, these authors found
that---with one exception---the $\Theta_0/R_0$ ratio should be
narrowly constrained between $\Theta_0/R_0 = 29.9$ km s$^{-1}$
kpc$^{-1}$ and $\Theta_0/R_0 = 31.6$ km s$^{-1}$ kpc$^{-1}$ (see also
Reid et al. 2009).

\begin{figure}[ht!]
\plotone{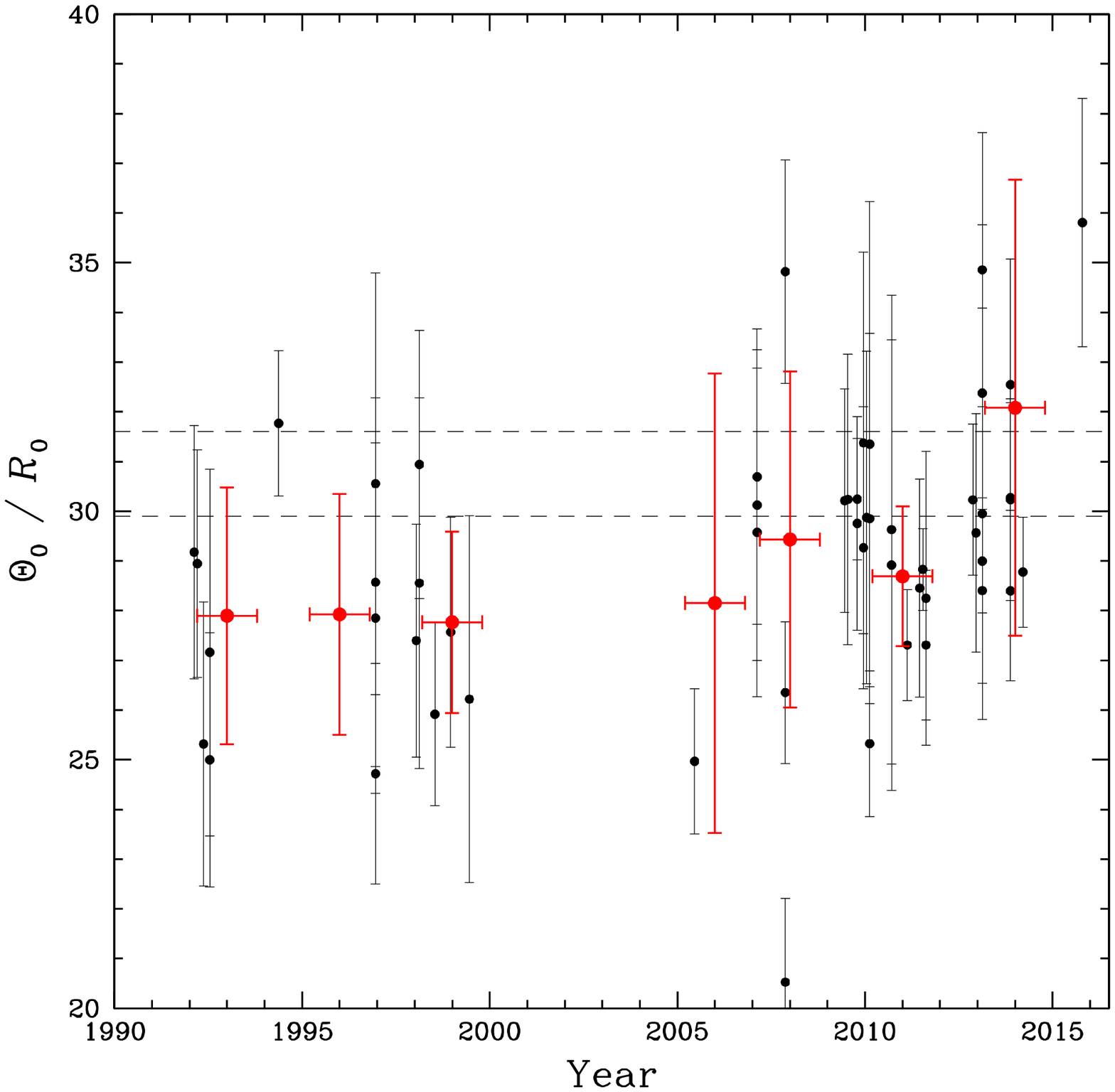}
\caption{$\Theta_0/R_0$ ratios used for determination of the Galactic
  Center distance based on kinematic disk modeling, published since
  1990. The horizontal dashed lines indicate the theoretically
  expected range (McMillan \& Binney 2010). The red data points are
  three-year averages, where the horizontal `error bars' indicate the
  relevant time period sampled and the vertical error bars represent
  the corresponding standard deviations.}
\label{fig4}
\end{figure}

However, Fig. \ref{fig4} shows that, in practice, the $\Theta_0/R_0$
ratios adopted by different authors are significantly smaller than the
range proposed by McMillan \& Binney (2010; see the horizontal dashed
lines in the figure). The red data points are three-year averages,
aimed at showing the general trend, where the horizontal `error bars'
indicate the relevant time period sampled and the vertical error bars
represent the corresponding standard deviations. Between 1990 and
2007, the majority of $\Theta_0/R_0$ ratios adopted by authors who
recently determined $R_0$ are found near $\langle \Theta_0/R_0 \rangle
= 28$ km s$^{-1}$ kpc$^{-1}$; since then, the ratio has slowly
increased to $\langle \Theta_0/R_0 \rangle \approx 32$ km s$^{-1}$
kpc$^{-1}$ (2013--2015).

Combined with the current-best Galactic Center distance estimates
based on S-star orbital modeling or statistical parallaxes, $R_0 = 8.3
\pm 0.2$ kpc, these $\Theta_0/R_0$ ratios imply that the Galactic
rotation speed at the solar circle is best constrained to the range
from $\Theta_0 = 232$ km s$^{-1}$ (1990--2007) to $\Theta_0 = 266$ km
s$^{-1}$ based on our current best understanding of Galactic dynamics,
i.e., significantly faster than the IAU-recommended value of
$\Theta_0^{\rm IAU} = 220$ km s$^{-1}$. In fact, this is confirmed
statistically: 33 of the 55 post-1990 values of $\Theta_0$ used in
this paper attained values of $\Theta_0 \ge 230$ km s$^{-1}$ (only 17
out of 55 publications derived $\Theta_0 < 220$ km s$^{-1}$). The
equivalent numbers for the period since 2000 are 25 out of 39 high
values (versus 10 out of 39 low rotation speeds). An increasing body
of evidence now suggests that the Galactic rotation speed at the solar
circle is indeed well in excess of the IAU recommendation (e.g., Reid
et al. 2009), with most authors supporting $\Theta_0 \in [235,260]$ km
s$^{-1}$ at the present time (e.g., Shen \& Zhu 2007; Bovy et
al. 2009; Zhu 2009; McMillan \& Binney 2010; Sato et al. 2010; Shen \&
Zhang 2010; Brunthaler et al. 2011; McMillan 2011; Honma et al. 2012;
Sch\"onrich 2012; Dambis et al. 2013; Reid 2013; Zhu \& Shen 2013;
Reid et al. 2014).

\section{Discussion}
\label{discussion.sec}

\subsection{Statistical trends}

Since the dual basic premises of science are reproducibility and
replicability (e.g., de Grijs 2016, and references therein), as
fallible humans scientists may be tempted---most likely
unknowingly---to trust results that are in good agreement with
previously published values. Such behavior may lead to a `bandwagon
effect,' more commonly referred to as publication bias (for
background, see Paper I).

In the context of Galactic Center distance determinations, the
possible presence of publication bias has been addressed repeatedly
(e.g., Reid 1989, 1993; Nikiforov 2004; Foster \& Cooper 2010; Malkin
2013a,b; Francis \& Anderson 2014), although their conclusions are not
in general agreement. Publication bias can be revealed either by a
gradual convergence of published values to a generally accepted number
(e.g., Malkin 2013a,b; Paper I) or by a sudden shift in preferred
values following a landmark publication.

Malkin (2013a,b) explored evidence of the former using his compilation
of 53 $R_0$ determinations published over a period of some 20 years
(1992--2011), but he did not find any statistically significant
trend.\footnote{At earlier times, the mean of the published Galactic
  Center distance estimates declined significantly and systematically,
  particularly in the period between approximately 1970 and 1990 (see
  Fig. \ref{fig1}; see also Reid 1989; 1993; Surdin 1999; Nikiforov
  2004), which may be the first suggestion of the presence of a
  bandwagon effect in $R_0$ estimates, although this could also simply
  be the result of improvements in either our physical understanding
  or of the methods used, or both.} Francis \& Anderson (2014) used an
unspecified, larger sample of 150 Galactic Center distances, cited as
``available online,'' of which they deemed 137 values as broadly
distinct, for a more detailed analysis of possible publication bias
affecting published Galactic Center distance estimates. These latter
authors claimed to have uncovered a continuing small but significant
trend toward decreasing estimates based on 48 values published between
1980 and 2000. (This encompasses approximately half the number of
distance determinations we recorded during the same period, i.e., our
database includes 100 Galactic Center distance determinations
published between 1980 February and 2000 September.) A Student's $t$
test implies that this trend's significance is reduced from 92\% prior
to 2000 to 73\% post-2000 (Francis \& Anderson 2014; see also Foster
\& Cooper 2010).

Note that even if the range of published values becomes narrower over
time, this does not necessarily imply that authors (or reviewers) are
unconsciously affected by a bandwagon effect. On the contrary, as we
already addressed in Paper I in the context of LMC distance
determinations, later publications may benefit from having access to
larger tracer samples than earlier papers, the calibration relations
are likely better understood, the analysis method applied are
improved, and systematic errors might be reduced (e.g., Malkin
2013a,b).

\subsection{Direct distance determinations}

Objective consideration of progress in the field of Galactic Center
distance determinations singles out one paper in particular that may
qualify as a landmark publication. Eisenhauer et al. (2003) were the
first to determine a direct, geometric distance to the Galactic Center
based on the orbital motion of S2, one of the small sample of stars
orbiting the Galactic Center's black hole on roughly Keplerian orbits.

Although the direct Galactic Center distance determinations by the
Genzel and Ghez groups are arguably the most precise Galactic Center
distance estimates available to date, their derivation is very
sensitive to a range of modeling assumptions (e.g., Gillessen et
al. 2009; Francis \& Anderson 2014). Two potentially important issues
the underlying assumption of Keplerian motion of the S stars around
Sgr A*. First, Newton's shell theorem, which posits that the gravity
inside a spherical shell is unimportant, provides only an approximate
solution to characterize the real, three-dimensional distribution of
matter in the nuclear disk and central bar of the Milky Way.

Second, the initial Keplerian approximations to the S-star orbits did
not take into account relativistic effects (Zucker et al. 2006;
Gillessen et al. 2009). Francis \& Anderson (2014) pointed out that
the corrections for the relativistic geodetic effect (de Sitter
precession) attempted by Gillessen et al. (2009) are based on the
assumption that this precession arises in the Schwarzschild metric,
whereas one should instead adopt the Kerr metric associated with a
relativistically rotating central black hole. In turn, this leads to
`frame dragging' (also known as the Lense--Thirring effect), which is
predicted to affect both the S stars' orbits and the light paths into
the line of sight. Neither Zucker et al. (2006) nor Gillessen et
al. (2009) considered these effects (for a detailed discussion, see
Francis \& Anderson 2014). Yet, in 2002, when the star S2 was near its
pericenter and relativistic effects were most significant, both Zucker
et al. (2006) and Gillessen et al. (2009) reported difficulties in
modeling the star's orbital motion. This thus implies that
relativistic effects must be taken into account and certainly cannot
be ignored.

Perhaps surprisingly, this appears to be a case where more data
actually lead to less accurate results. The potentially significant
relativistic effects affecting the S-star orbits result in ever more
serious cumulative differences from Keplerian motion (Francis \&
Anderson 2014). Nevertheless, here we consider the $R_0$
determinations based on the orbital motions of the S stars as a
function of publication date. We refer to Fig. \ref{fig1}b and Table
\ref{GCdirect.tab} (top) for guidance. The most significant change in
the central $R_0$ values since Eisenhauer et al. (2003, 2005) occurred
with the publication of Gillessen et al. (2009), which resulted in an
upward adjustment of approximately 0.6 kpc.

Gillessen et al. (2009) point out that their new values for $R_0$ are
consistent within the errors with the Eisenhauer et al. (2003, 2005)
results, and that adding more stars to the orbital solutions does not
change the results significantly. They assert that their main
improvement is owing to the more rigorous treatment of the systematic
errors. In fact, it appears that this change is indeed not driven by
publication bias, nor did their consideration of relativistic effects
have a significant impact. The latter statement is supported by
comparing the Zucker et al. (2006) Galactic Center distance estimate
with the earlier determinations of Eisenhauer et al. (2003,
2005). Instead, the Gillessen et al. (2009) Galactic Center distance
is more robust than their previous results, given that it is based on
astrometric observations of much higher quality (affected by smaller
astrometric uncertainties) and covering a significantly longer time
span.

The latest results obtained by both Genzel's group (Gillessen et
al. 2009, 2013) and Ghez' collaboration (Ghez et al. 2008; Yelda et
al. 2011) are indeed fully mutually consistent, as also expected from
the $R_0$ values resulting from combining the data of both groups
(Gillessen et al. 2009). Indeed, the low value of $R_0 = 7.7 \pm 0.4$
kpc derived by Morris et al. (2012), members of the Ghez group, is an
odd one out. These latter authors only commented that this estimate
corresponds to their current best fit to the observational data from
1995--2011. They did not discuss their discrepant result, other than
by pointing out that the uncertainty in their estimate is highly
correlated with the uncertainty in the mass of the black hole. Indeed,
the same degeneracy affects the central values (e.g., Gillessen et
al. 2009). In conclusion, we have not unearthed any evidence of
possible publication bias in the Galactic Center distances resulting
from S-star orbital modeling.

Careful assessment of the $R_0$ values obtained from statistical
parallax measurements of the nuclear star cluster also show broad
consistency during the 20 years leading up to 2015 (for a brief
discussion, see Section \ref{direct.sec}). Therefore, the current-best
distance determination to the Galactic Center is, arguably, based on
appropriate combined direct observations of both the S-star orbits and
the Galactic Center's statistical parallax properties. Do et
al. (2013) and Chatzopoulos et al. (2015) both attempted such a
combination, yielding $R_0 = 8.46^{+0.42}_{-0.38}$ kpc and $R_0 = 8.33
\pm 0.11$ kpc, respectively.\footnote{Both studies were, in essence,
  modern applications of the idea to combine velocity dispersions with
  proper motion measurements in the Galactic bulge, first proposed by
  Minniti (1993) and Kuijken (1995).} However, other recent direct
Galactic Center distance measurements yield somewhat smaller distances
(e.g., Trippe et al. 2008, 2011; Yelda et al. 2011; Gillessen et
al. 2013). At the present time, we therefore advocate a benchmark
Galactic Center distance of $R_0 = 8.3 \pm 0.2$ kpc, where the
uncertainty is the statistical error resulting from combination of the
Do et al. (2013) and Chatzopoulos et al. (2015) `combined' values. The
additional, systematic uncertainty, which should be added in
quadrature, is of order $\pm 0.4$ kpc. The latter estimate follows
from consideration of the full set of updated values of $R_0$ in Table
\ref{GCdirect.tab}, where the variety of assumptions made is reflected
in the spread of the central values (see also the discussion in
Section \ref{direct.sec}).

\subsection{Indirect distance determinations}

The unprecedented volume of our catalog of Galactic Center distance
determinations has allowed us to conclude that the globular cluster,
Cepheid, and red clump centroids suggest smaller Galactic Center
distances than the direct distances estimates, while the RR Lyrae and
Mira variables result in larger Galactic Center centroid distances
(see, e.g., Table \ref{GCcentroids.tab}).

Indeed, it has been suggested that some tracers, and in particular the
($t \ge 10$ Gyr) old tracers (i.e., the RR Lyrae stars) might or might
not follow the Galactic Bar (e.g., Zoccali \& Valenti 2016). This
implies that distance determinations based on these tracers might
depend on the line of sight of the adopted sample. Moreover, we do not
yet have similarly robust identifications of RR Lyrae stars located in
the Galactic Center center as we do for the Galactic Center Cepheid
sample. This suggests that RR Lyrae variables might be more prone to
possible systematic effects (e.g., Zoccali \& Valenti 2016). In
addition, there are reasons to believe that the red clump results
might also be affected by large systematic uncertainties. In fact, if
a metallicity gradient is present from the Galactic Center to the
outer bulge (e.g., Zoccali et al. 2008; Pietrukowicz et al. 2012,
2015; Uttenthaler et al. 2012), this also means that the mean
magnitude and color of red clump stars change when approaching the
Galaxy's inner core.

A separate concern regarding systematic effects introduced by our
homogenization is the following. If, as we have done for many RR
Lyrae-based Galactic Center distance determinations, we adopt any of
the prevailing $M_V$ versus [Fe/H] relations, we have to deal with a
strong dependence on the reddening correction and the metallicity
distribution of RR Lyrae in the central Galaxy. Specifically, we have
adopted $M_V{\rm (RR)} = 0.72$ mag. If the original papers quoting
Galactic Center distance determinations used any other absolute
magnitude calibration (see Table \ref{GCcentroids.tab}), we converted
the $R_0$ value to a distance modulus, corrected it for the difference
in calibration used, and eventually converted it back to a linear
distance. If we were to adopt the $K$-band magnitudes for our
calibration instead, the dependence on metallicity and on the
reddening uncertainties is smaller. However, in this case we would
introduce a dependence on the pulsation period, since RR Lyrae
exhibiting longer periods are brighter. Where appropriate, we have
also assumed that ${\rm [Fe/H]} = -1.0$ dex for RR Lyrae in the bulge
(e.g., Bono et al. 2003). If we assume that a typical mean period for
a fundamental RR Lyrae in the bulge is around 0.55--0.60 days and the
mean metallicity is around [Fe/H]$ = -1$, $\langle M_K \rangle = -0.5$
mag (Marconi et al. 2015). At the present time, we have access to too
few $K$-band calibrated RR Lyrae distance measurements to the Galactic
Center (only two such calibrations are available: Carney et al. 1995;
Dambis 2009). Overall, the adjustments thus made take into account the
metallicity dependence of the RR Lyrae distance scale to the best of
our current ability.

Finally, we explored whether any of the centroid tacers or those used
for the kinematics-based Galactic Center distance estimates exhibited
statistically significant trends in the modern, post-1990 era. In the
context of exploring possible bandwagon effects, one should, of
course, only consider the {\it original} Galactic Center distance
estimates, before homogenization. The significant scatter among the
data points pertaining to any of the kinematics tracers precludes us
from resolving any such trend. Among the centroid tracers, we do not
have sufficient numbers of data points for the globular clusters,
planetary nebulae, or Cepheids spanning a sufficiently long time span
to draw statistically justifiable conclusions either (see
Fig. \ref{fig2}).

For all other tracers we only find statistically {\it insignificant}
($\la 1 \sigma$) trends of $R_0$ as a function of publication date,
both for the full 1990--2016 time span and for the more recent period
since 2005. However, we find that in general the uncertainties have
been steadily decreasing for all tracers, which largely owing to
improvements in the calibration approaches used and access to new and
larger samples of tracer objects (for a related discussion, see Paper
I). In conclusion, neither the centroid- nor the kinematics-based
Galactic Center distance tracers published since 1990 suggest the
presence of any significant trends of $R_0$ with publication date.

\section{Summary and Conclusions}

Aiming at deriving a statistically well-justified Galactic Center
distance based on a large variety of tracers and reducing any
occurrence of publication bias, we embarked on an extensive
data-mining effort of the scientific literature, eventually yielding
273 new or revised $R_0$ estimates published since records began in
October 1918 until June 2016. Our large database of Galactic Center
distance estimates, a fully linked version of which is made available
to the scientific community, allowed us to explore the pros and cons
of a variety of different approaches used to determine the distance to
the Galactic Center.

We separated our compilation into direct and indirect distance
measurements. The former include distances such as those based on
orbital modeling of the so-called S stars orbiting Sgr A*, the closest
visual counterpart of the Milky Way's central supermassive black hole,
as well as those relying on statistical parallaxes of either the
nuclear star cluster or the stellar population in the inner Galactic
core. Careful assessment of the body of published Galactic Center
distances based on these methods resulted in our Galactic Center
distance recommendation of $R_0 = 8.3 \pm 0.2 \mbox{ (statistical)}
\pm 0.4 \mbox{ (systematic)}$ kpc.

A much larger body of Galactic Center distance determinations is based
on indirect methods, either those relying on centroid determinations
for a range of different tracer populations (e.g., globular clusters,
Cepheid, RR Lyrae, or Mira variables, or red clump stars) or
measurements based on kinematic observations of objects at the solar
circle, combined with a mass and/or rotational model of the Milky
Way. The latter approaches are affected by significantly larger
uncertainties than the former, while the central, mean Galactic Center
distances based on the kinematic methods are systematically smaller
than those based on centroid determinations. Most centroid-based
distances are in good agreement with those resulting from the direct
methods.

We did not find any conclusive evidence of the presence of a bandwagon
effect in the post-1990 Galactic Center distance measurements, neither
among the direct distance estimates nor among those based on centroid
determinations. Our set of kinematics-based distance measurements
cannot be used to explore this issue given the significant
uncertainties associated with the latter methods. However, these
latter measurement can indeed be used to constrain the Galaxy's
rotation velocity at the solar Galactocentric distance using the
$\Theta_0/R_0$ ratios employed by their respective authors. We found a
gradual increase in the mean value of $\Theta_0$ from $\langle
\Theta_0 \rangle = 232$ km s$^{-1}$ in the 1990s to $\langle \Theta_0
\rangle = 266$ km s$^{-1}$ more recently. (Both values represent the
means of the distributions of rotation speeds; their standard
deviations are of order 20--30 km s$^{-1}$.) Our results thus imply
that the IAU-recommended Galactic Center distance ($R_0^{\rm IAU} =
8.5$ kpc) needs a downward adjustment, while the recommendation for
the Galactic rotation velocity at the solar circle ($\Theta_0 = 220$
km s$^{-1}$) requires an upward revision.

Finally, in view of the recent {\sl Gaia} Data Release 1 (Lindegren et
al. 2016), this is an opportune time to consider the impact on
Galactic Center distance determinations of the improved parallax
measurements of upwards of a billion stars that the mission will
provide by the end of its nominal five-year duration. The final {\sl
  Gaia} catalog will map a significant fraction of the Galactic
volume, with particularly high (distance-dependent) parallax precision
for stars in the solar neighborhood, gradually decreasing toward the
Galactic Center.

Because of the high extinction and significant crowding in the
Galaxy's inner regions, {\sl Gaia}'s optical ($G$-band) measurements
will reach the Galaxy's central regions, but not the Center itself. It
is anticipated that {\sl Gaia}'s direct distance determinations of
stars near the Galactic Center will be accurate to approximately 20\%,
i.e., significantly worse than the distance estimates provided by any
of the direct methods discussed in Section \ref{direct.sec} or the
Cepheid-based result of Matsunaga et al. (2011). However, {\sl Gaia's}
homogeneous and extensive final catalog will undoubtedly facilitate a
significantly improved {\it statistical} determination of $R_0$ based
on kinematic measurements at as well as inside the solar circle. In
addition, {\sl Gaia} will provide improved distance measurements to
large numbers of standard candles, thus almost certainly improving
their zero-point accuracies, which in turn will allow us to test for
the effects of population differences associated with the use of
secondary distance tracers.

{\sl Gaia} is currently among the best-placed facilities to resolve
the remaining uncertainties in the Galactic distance scale, which is
no longer seriously affected by statistical uncertainties. The
prevailing uncertainties preventing us from determining a more
accurate distance to the Galactic Center are systematic in
nature. Among the latter, the effects of not just variations in the
extinction, but of possible variations in the prevailing extinction
{\it law} are among the most important stumbling blocks.

The data set analyzed in this paper is, unfortunately, unsuitable for
systematic studies of the uncertainties introduced by variations in
the extinction law (e.g., Nishiyama et al. 2009; Nataf et al. 2016;
and references therein). This is particularly so, because the vast
majority of the individual data points retrieved from the literature
have been extinction-corrected by their respective authors using the
most appropriate approaches available at the time of their
analyses. This implies the inherent presence of intrinsic
inhomogeneities in the extinction corrections, which are impossible to
fully correct for at the present time.

In addition, we point out that current empirical estimates of
reddening laws in the local Universe are mainly based on optical and
near-infrared photometry. To further constrain possible systematic
effects, independent approaches based on either analyses of diffuse
interstellar bands (e.g., Munari \& Zwitter 1997; Wallerstein et
al. 2007; Munari et al. 2008; Kashuba et al. 2016) or spectroscopic
studies (e.g., Kudritzki et al. 2012) may shed new light on this
long-standing problem. Indeed, detailed, large-scale studies such as
that of Nataf et al. (2016), perhaps combined with high spatial
resolution observations of carefully selected, homogeneous samples of
Galactic Center objects at near- to mid-infrared wavelengths (such as
those anticipated to result from {\sl WFIRST} operation or from
campaigns with the next generation of 30 m-class ground-based
telescopes), seem most promising to overcome the remaining systematics
in the reddening laws.

\section*{Acknowledgements}

R. d. G. is grateful for research support from the National Natural
Science Foundation of China through grants 11373010, 11633005, and
U1631102. This work was also partially supported by PRIN-MIUR
(2010LY5N2T), `Chemical and dynamical evolution of the Milky Way and
Local Group galaxies' (PI F. Matteucci). This research has made
extensive use of NASA's Astrophysics Data System Abstract Service.

\end{document}